\documentclass[twocolumn,showpacs,preprintnumbers,amsmath,amssymb,superscriptaddress,showkeys,prd]{revtex4}

\usepackage{graphicx}
\usepackage{dcolumn}
\usepackage{bm}
\usepackage{amsmath,amssymb}

\begin{document}

\preprint{\emph{Revista Mexicana de Astronom\'{\i}a y
Astrof\'{\i}sica}}

\title{Temperature, brightness and spectral index of the Cygnus radio loop}

\author{V. Borka Jovanovi\'{c}}
\email[Corresponding author:]{vborka@vinca.rs} \affiliation{Atomic
Physics Laboratory (040), Vin\v{c}a Institute of Nuclear Sciences,
University of Belgrade, P.O. Box 522, 11001 Belgrade, Serbia}

\author{D. Uro\v{s}evi\'{c}}
\affiliation{Department of Astronomy, Faculty of Mathematics,
University of Belgrade, Studentski trg 16, 11000 Belgrade, Serbia}

\date{March 11, 2011}

\begin{abstract}
The estimated brightness of the Cygnus loop supernova remnant (SNR)
at 2720, 1420, 820, 408 and 34.5 MHz are presented. The observations
of the continuum radio emission are used to calculate the mean
brightness temperatures and surface brightnesses of this loop at the
five frequencies in wide spectral range, using the method we have
previously developed for large radio loops. The spectrum for mean
temperatures versus frequency between the five frequencies is
estimated and the spectral index of Cygnus loop is also obtained.
Also, from our results can be concluded that Cygnus loop evolves in
the low density environment and the initial energy of supernova
explosion was relatively low. The obtained results confirm
non-thermal origin of the Cygnus radio loop and show that our method
is applicable to almost all remnants.
\end{abstract}



\maketitle

\section{Introduction}
\label{sec01}

In our investigation, we are supposing the radio loops to be evolved
supernova remnants (SNRs). Supernova remnants represent shells of
approximately spherical shape which are spreading, and depending of
the interstellar matter density, they change their shape. We can see
only parts of these spherical contours which represent radio spurs,
e.g. areas of more intense emission in the sky, spur-like and of
huge dimensions. Radio loops consist of more spurs lying
approximately in the same small circle of the celestial sphere.
Their material probably expands inside of bubbles of low density.
Bubbles are made by former SNR explosions or by strong stellar winds
(\citet{salt83,mcke77} and references therein). The radio emission
from SNRs is generally understood to be synchrotron emission from
the relativistic electrons moving in the magnetic field.

A star in the constellation of Cygnus exploded and its remnant is
Cygnus loop. It is classified as a middle-aged SNR of type S,
located below (but near the plane of) the Galactic equator, and less
than 1 kpc away from us. It is listed in Green's catalogue of SNRs
as G74.0-8.5 (\citet{gree04,gree06,gree09}). As shown in
\citet{gree84}, this remnant has been decelerated considerably by
its interaction with the surrounding interstellar medium.

\citet{leah97} presented high resolution 1420 MHz total intensity
and polarization maps of the Cygnus loop and derived rotation
measures. \citet{asch99} made a comparison between the radio and
X-ray emission from supernova remnants. The X-ray emission traces
the hot gas behind the shock front, while the radio emission comes
from the relativistic electron population emitting synchrotron
radiation in the ambient magnetic field. They found that there are
significant differences in the distribution of X-ray and radio
brightness in the Cygnus loop. Also, significant temperature
variations, seen along the rim and in the bright filaments interior
to the rim, pointed out that the Cygnus loop was caused by a
supernova exploding in a cavity.

\citet{uyan02} suggested that Cygnus loop may consist of two
overlapping remnants. Substantial differences between the northern
and southern part as well as in their emission characteristics have
been observed (\citet{uyan04,patn02}). Some authors proposed that
the Cygnus loop consists of two likely interacting SNRs: G74.3-8.4
and G72.9-9.0 (\citet{uyan02,leah02}). In \citet{sun06}, from 4800
MHz observations, it is explained that the polarization maps (the
difference in the polarization characteristic between the northern
and southern part) support previous ideas that the Cygnus loop may
consist of two SNRs. Besides, several compact radio sources are
located within the boundary of the remnant. The main characteristics
of the Cygnus loop in different spectral bands are: (a) in radio
band: it is shell, brightest to the north-east, with fainter
breakout region to south, with spectral variations, (b) optical:
large filamentary loop, brightest to the north-east, not well
defined to the south and west, (c) X-ray: shell in soft X-rays (see
e.g. in \citet{gree06}).

Some radio maps which include this remnant are the following: the
map by \citet{kund72} at 4940 MHz, \citet{keen73} at 2695 MHz,
\citet{uyan04} at 2675 MHz, \citet{leah97} at 1420 MHz,
\citet{dick80} at 610 MHz, \citet{gree84} at 408 MHz. Observations
of the continuum radio emission at 2720 (\citet{reif87}), 1420
(\citet{reic86}), 820 (\citet{berk72}), 408 (\citet{hasl82}) and
34.5 MHz (\citet{dwar90}), we found in electronic form and used them
in this paper.

In our previous paper (\citet{bork09a}) we only calculated the
temperatures and brightnesses at the three frequencies. In this
paper we expand the scope of our investigation to: brightness
temperatures and surface brightnesses at the five frequencies, as
well as spectrum, $T-T$ graphs, the radio spectral indices,
estimation of environment density and the initial energy of
supernova explosion, and the flux density spectrum.

In our calculations, we used brightness temperatures over the whole
area of the loop, so the mean temperature that we estimated refers
to northern and southern parts together. In this research the
average brightness temperatures and surface brightnesses of the
Cygnus radio loop are calculated at the five frequencies: 2720,
1420, 820, 408 and 34.5 MHz. Then we study how these results are
getting along with previous results (\citet{roge99,reic03,uyan04})
and with current theories of SNR evolution. These theories predict
that loops are non-thermal sources which are spreading inside of the
hot and low density bubbles made by former supernova explosions or
by strong stellar winds (see \citet{salt83,mcke77} and references
therein).

Our aim is also to apply method for determination of the brightness
temperature given in article \citet{bork07} which is developed for
large radio loops, and to show that it is rather efficient in the
case of much smaller radio loops, e.g. Cygnus loop, as well as to
check how the results obtained using this method are getting along
with the method of $T-T$ graphics and results obtained with other
methods. Our method is quite simple because we are using brightness
temperature isolines to define borders of the Cygnus loop at the
wide range of frequencies ($\nu_{max} / \nu_{min}$ = 2720 MHz / 34.5
MHz $\approx$ 79). Other authors are using different squared or
rectangular areas to determine area of the loop and calculate
spectral indices, brightness temperature and the flux of the loop
(\citet{uyan04}, page 917 and \citet{leah98}, page 786). Also, we
calculated flux from Cygnus loop and compare our results with
results of other authors in different frequency ranges.

\section{Analysis}
\label{sec02}

\subsection*{Data}
\label{sec02a}

\begin{figure}[ht!]
\centering
\includegraphics[width=0.45\textwidth]{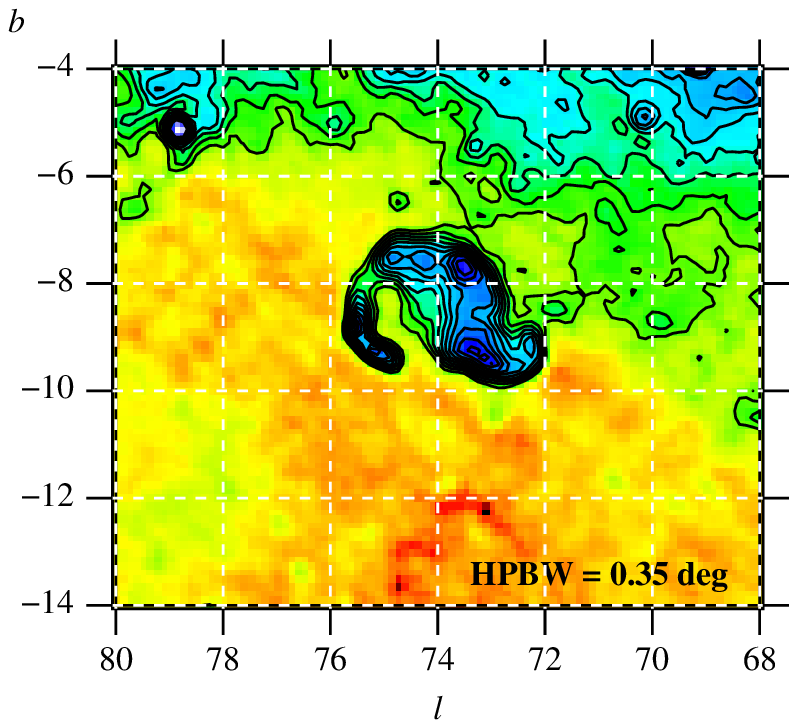} \\
\vspace*{0.2cm}
\hspace*{0.4cm}
\includegraphics[width=0.34\textwidth]{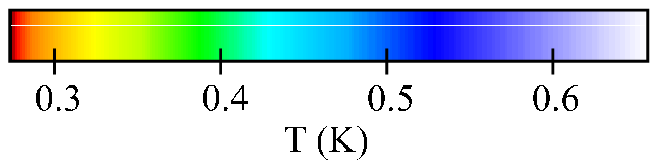} \\
\vspace*{0.8cm}
\includegraphics[width=0.50\textwidth]{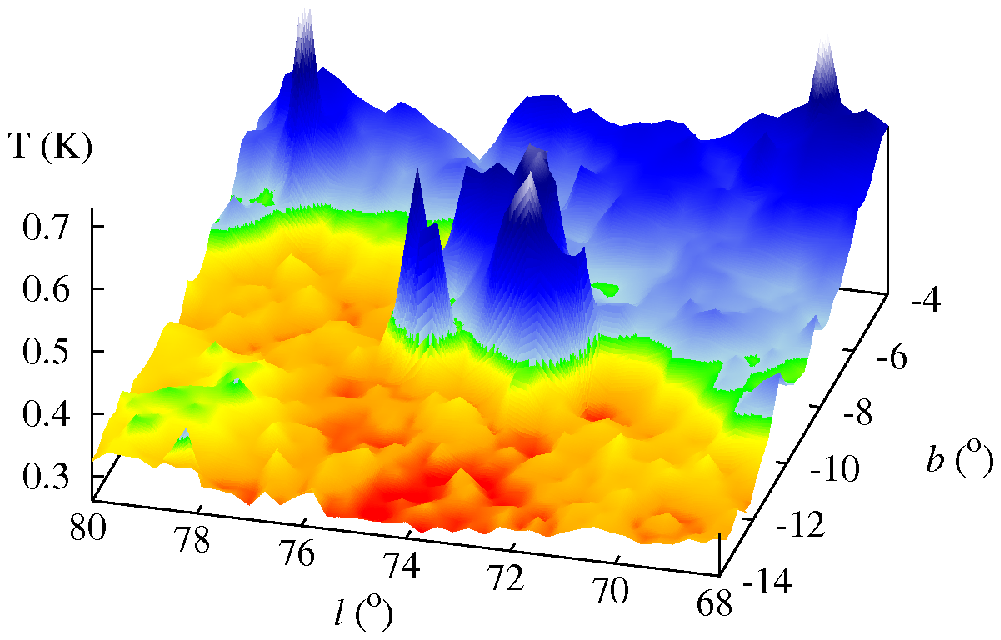}
\caption{\emph{Top}: the 2720 MHz map of a region in Cygnus, in new
Galactic coordinates ($l$, $b$), showing contours of brightness
temperature. This radio loop has position: $l$ = [76.5$^\circ$,
71.5$^\circ$]; $b$ = [-10.5$^\circ$, -7$^\circ$]. The HPBW (Half
Power Beam Width) for this frequency is 0$^\circ$.35. Eleven
contours plotted represent the temperatures $T_\mathrm{min}$ and
$T_\mathrm{max}$ from Table \ref{tab01} and nine contours in
between. The contours are plotted every 0.265 K, starting from the
lowest temperature of 0.395 K up to 0.66 K. The corresponding
temperature scale is given (in K). \emph{Bottom}: the 2720 MHz area
map of Cygnus.}
\label{fig01}
\end{figure}

\begin{figure}[ht!]
\centering
\includegraphics[width=0.45\textwidth]{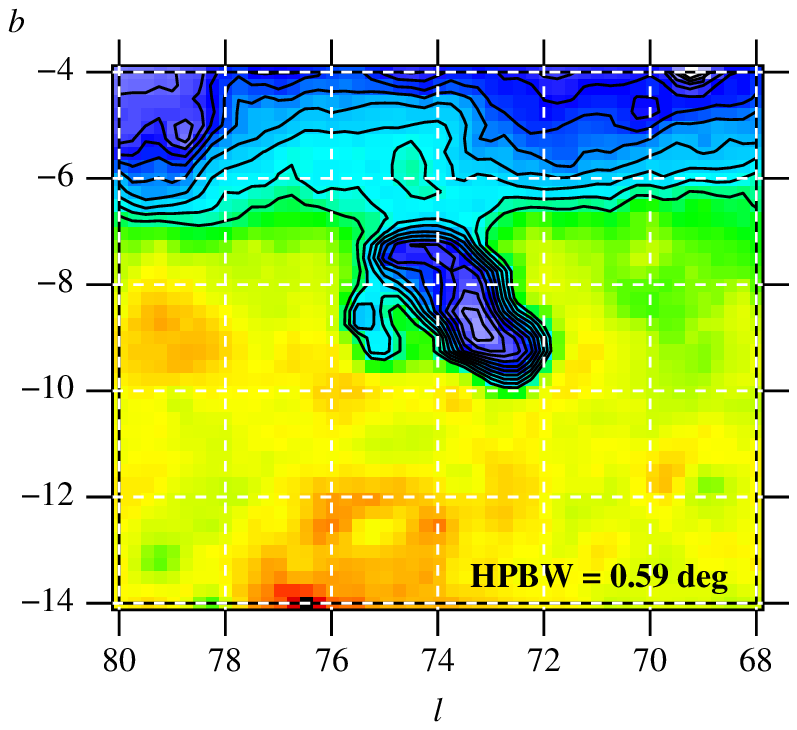} \\
\vspace*{0.2cm}
\hspace*{0.4cm}
\includegraphics[width=0.34\textwidth]{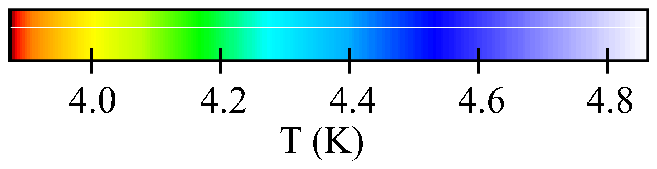} \\
\vspace*{0.8cm}
\includegraphics[width=0.50\textwidth]{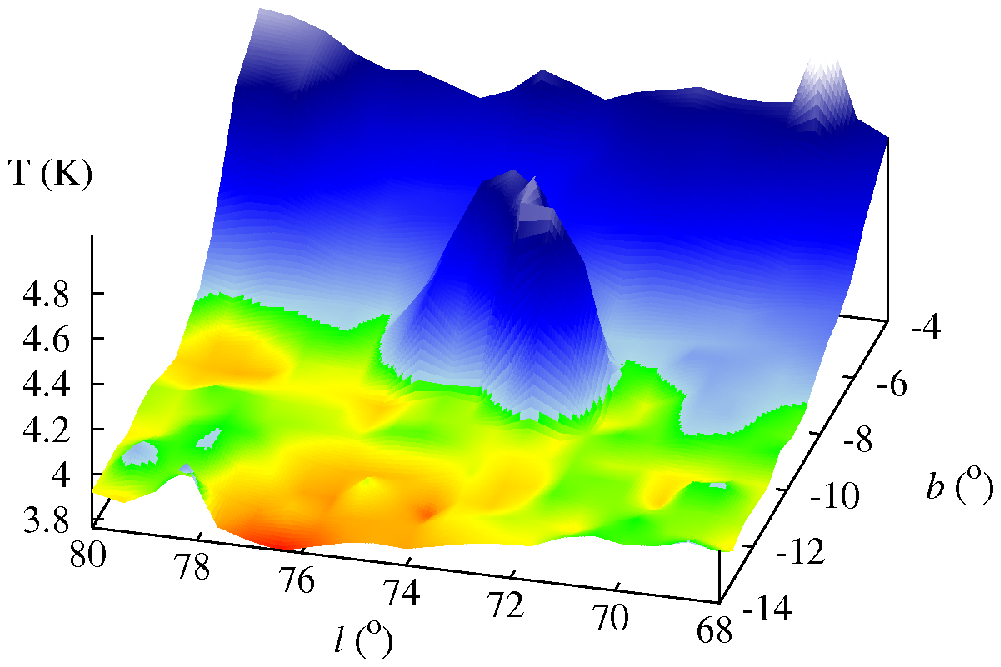}
\caption{The same as Fig. 1, but for 1420 MHz. The HPBW for this
frequency is 0$^\circ$.59. The contours are plotted every 0.07 K,
starting from the lowest temperature of 4.2 K up to 4.9 K.}
 \label{fig02}
\end{figure}

\begin{figure}[ht!]
\centering
\includegraphics[width=0.45\textwidth]{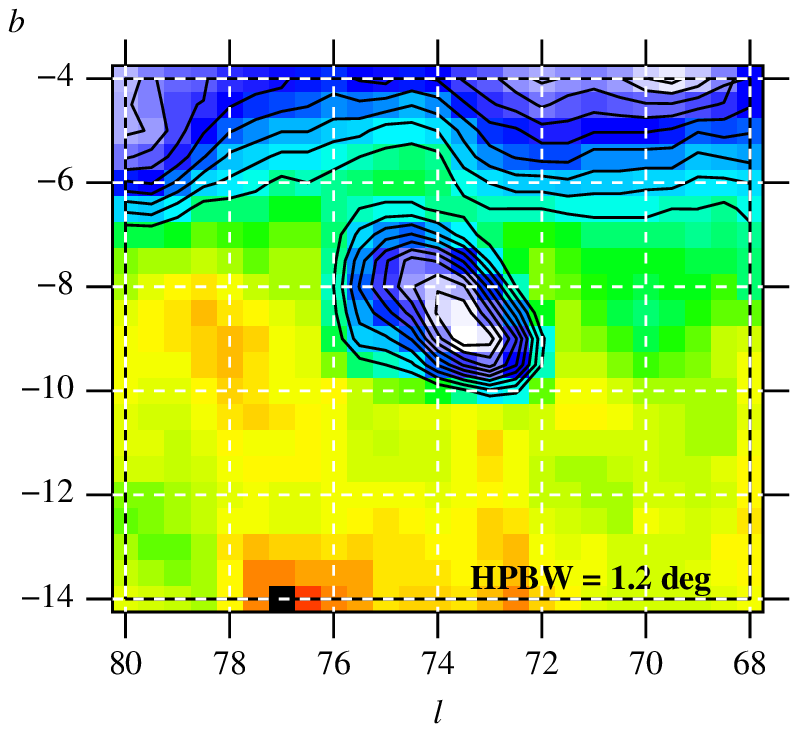} \\
\vspace*{0.2cm}
\hspace*{0.4cm}
\includegraphics[width=0.34\textwidth]{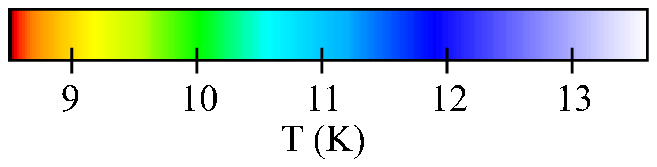} \\
\vspace*{0.8cm}
\includegraphics[width=0.50\textwidth]{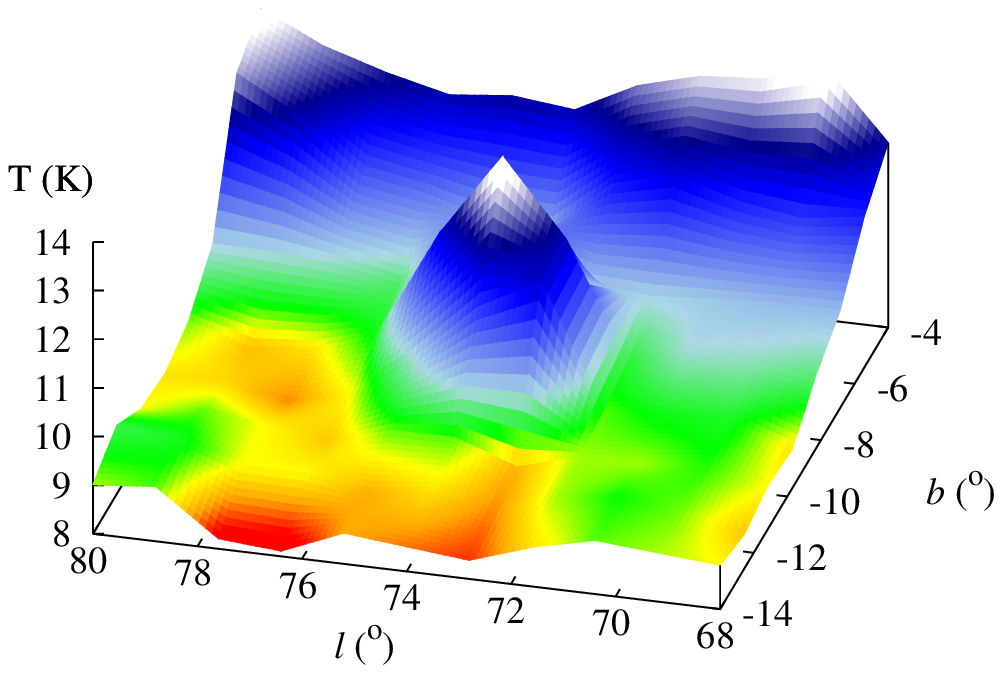}
\caption{The same as Fig. 1, but for 820 MHz. The HPBW is
1$^\circ$.2. The contours are plotted every 0.39 K, starting from
the lowest temperature of 10.1 K up to 14 K.}
\label{fig03}
\end{figure}

\begin{figure}[ht!]
\centering
\includegraphics[width=0.45\textwidth]{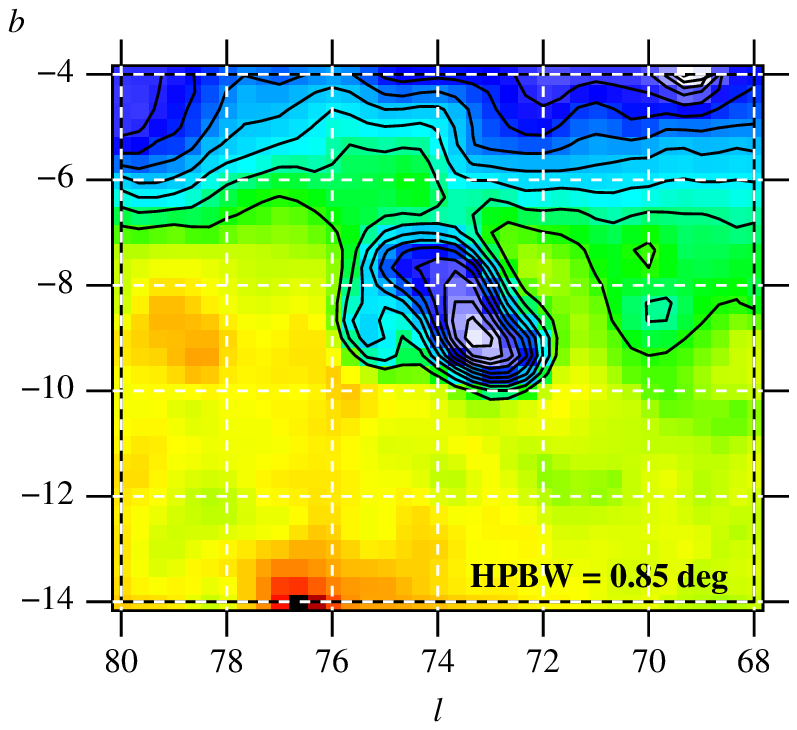} \\
\vspace*{0.2cm} \hspace*{0.4cm}
\includegraphics[width=0.34\textwidth]{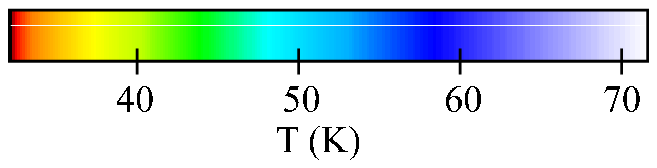} \\
\vspace*{0.8cm}
\includegraphics[width=0.50\textwidth]{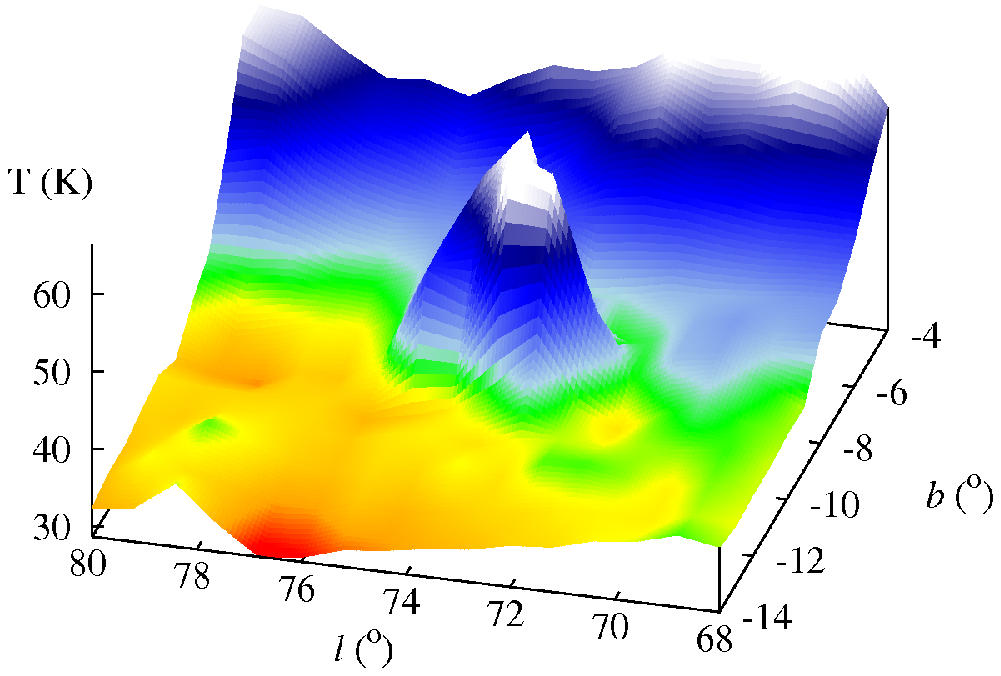}
\caption{The same as Fig. 1, but for 408 MHz. The HPBW is
0$^\circ$.85. The contours are plotted every 3.1 K, starting from
the lowest temperature of 41 K up to 72 K.}
\label{fig04}
\end{figure}

\begin{figure}[ht!]
\centering
\includegraphics[width=0.45\textwidth]{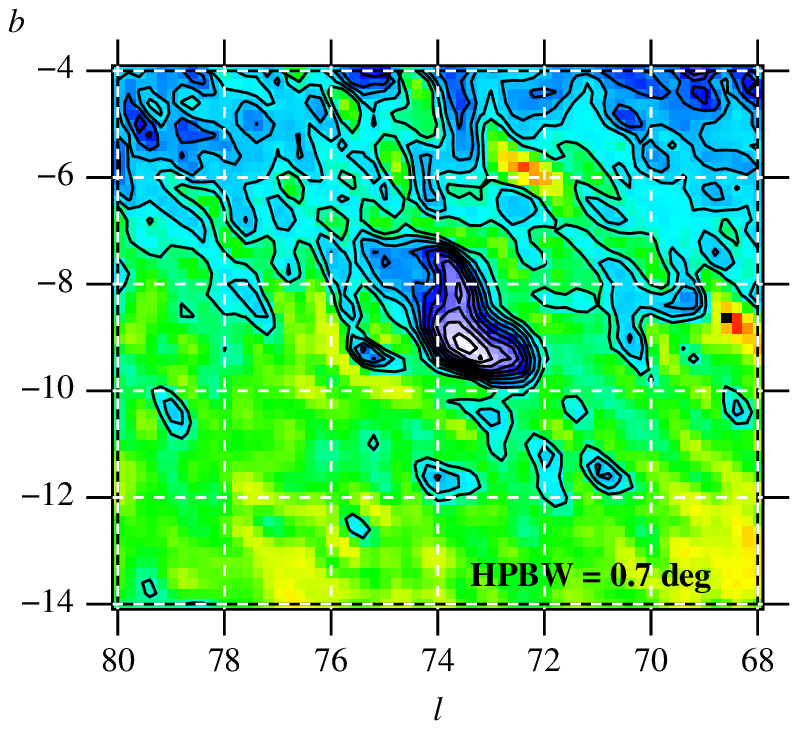} \\
\vspace*{0.2cm} \hspace*{0.4cm}
\includegraphics[width=0.34\textwidth]{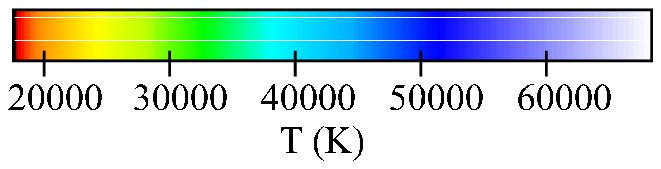} \\
\vspace*{0.8cm}
\includegraphics[width=0.50\textwidth]{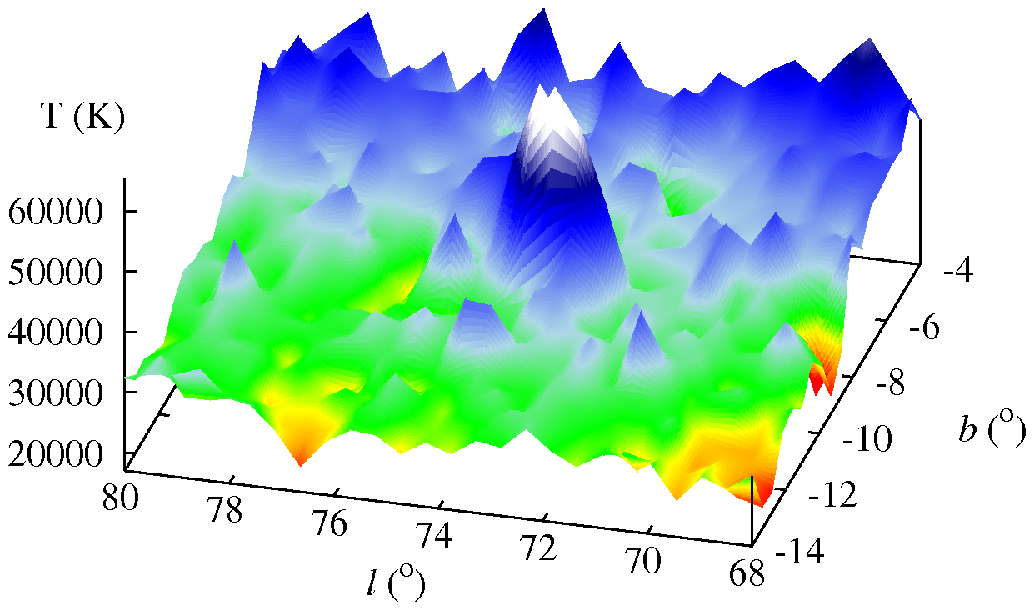}
\caption{The same as Fig. 1, but for 34.5 MHz. The HPBW is
0$^\circ$.7. The contours are plotted every 3550 K, starting from
the lowest temperature of 34500 K up to 70000 K.}
\label{fig05}
\end{figure}

\begin{figure*}[ht!]
\centering
\includegraphics[height=0.9\textheight]{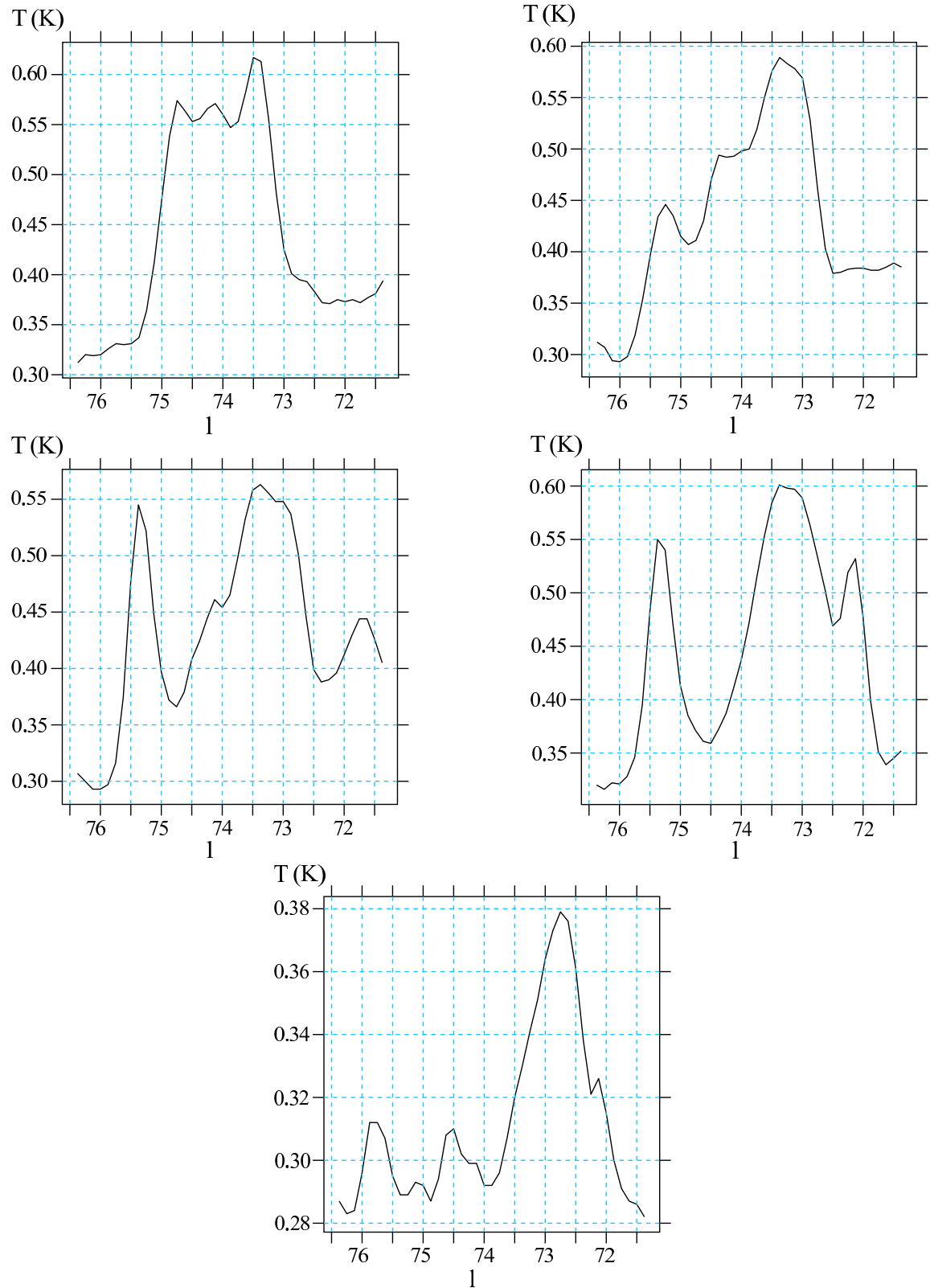}
\caption{Temperature profiles at 2720 MHz for Galactic longitude
from 76$^\circ$.5 to 71$^\circ$.5 and for the following values of
Galactic latitude: $b$ = -7$^\circ$.5 \emph{(top left)}, $b$ =
-8$^\circ$ \emph{(top right)}, $b$ = -8$^\circ$.5 \emph{(middle
left)}, $b$ = -9$^\circ$ \emph{(middle right)} and $b$ = -10$^\circ$
\emph{(bottom)}. Temperatures are given in K, and Galactic
longitudes in degrees. Notice that the bottom figure shows the
temperature profile at latitude which lies outside (but close enough
to) the loop and from that profile we can see what is the highest
temperature of the background at $b$ = -10$^\circ$, at 2720 MHz (not
of the loop with background).}
\label{fig06}
\end{figure*}

We used the following radio-continuum surveys as the basic source of
data in this paper: at 2720 MHz (\citet{reif87}), 1420 MHz
(\citet{reic86}), 820 MHz (\citet{berk72}), 408 MHz (\citet{hasl82})
and 34.5 MHz (\citet{dwar90}). These surveys are available in
electronic form, as the "Flexible Image Transport System" (FITS)
data format, at the site:
\url{http://www.mpifr-bonn.mpg.de/survey.html}. This online Survey
Sampler of the "Max-Planck-Institut f\"{u}r Radioastronomie" (MPIfR)
near Bonn, Germany, allows users to pick a region of the sky and
obtain images and data at different frequencies. The 2720-MHz
Stockert survey has the angular resolution of 0$^\circ$.35, 1420-MHz
Stockert survey (\citet{reic86}) 0$^\circ$.59, the 820-MHz Dwingeloo
survey (\citet{berk72}) 1$^\circ$.2, the 408-MHz all-sky survey
(\citet{hasl82}) 0$^\circ$.85 and the 34.5-MHz Gauribidanur survey
(\citet{dwar90}) 0$^\circ$.7. The corresponding observations are
given at the following rates (measured data) for both $l$ and $b$:
$\frac{1^\circ}{8}$ at 2720 MHz, $\frac{1^\circ}{4}$ at 1420 MHz,
$\frac{1^\circ}{2}$ at 820 MHz, $\frac{1^\circ}{3}$ at 408 MHz and
$\frac{1^\circ}{5}$ at 34.5 MHz. The effective sensitivities are
about 5 mK T$_b$ (T$_b$ is for an average brightness temperature),
50 mK, 0.20 K, 1.0 K and about 700 K, respectively.

FITS data format stores the multidimensional arrays and
2-dimensional tables containing rows and columns of scientific data
sets. We extracted observed brightness temperatures from this data
format into ASCII data files, and afterwards using our programs in C
and FORTRAN, we obtained results presented in this paper.

\subsection*{Method}
\label{sec02b}

As there is great influence of background radiation over the Cygnus
loop area, it is very difficult to determine its area precisely. The
area of this loop is enclosed with brightness temperature contours.
The maps of a region in Cygnus, in new Galactic coordinates ($l$,
$b$), with contours of the brightness temperatures $T_\mathrm{b}$
are plotted in Figs. \ref{fig01}--\ref{fig05}. In these figures,
among all the brightness temperature contours, the most important
are the outer contour and inner one which represent the loop
borders. The outer one (which corresponds to the minimum temperature
of the loop) separates loop and background, while the inner one
(corresponding to the maximum temperature) separates loop and some
superposed source. The interval of Galactic longitude of this loop
is $l$ = [76.5$^\circ$, 71.5$^\circ$], and of latitude is $b$ =
[-10.5$^\circ$, -7$^\circ$]. We used the same method of calculation
as given in \citet{bork07} for Galactic radio loops I--VI,
\citet{bork08} for Loops V and VI and in \citet{bork09b} for
Monoceros loop. Our aim is to apply method, which we developed for
main Galactic loops I-VI and described in \citet{bork07} and
\citet{bork08}, to smaller remnants and to show that our method of
calculation is applicable to almost all remnants. To confirm that
our method is good, we also gave the area maps of Cygnus at the five
frequencies (Figs. \ref{fig01}--\ref{fig05}) and some examples of
the temperature profiles at the 2720 MHz (Fig. \ref{fig06}). In Fig.
\ref{fig06} it should be noticed that the bottom panel shows only
background radiation (not including loop), where for $b$ =
-10$^\circ$ the brightness temperature is not higher than 0.38 K. It
is in agreement with temperature intervals given in the first row of
the Table \ref{tab01}, where for the minimum temperature of the loop
with background, at 2720 MHz, it is put 0.395 K.

\clearpage

The mean temperatures and surface brightnesses of this radio loop
are computed using data taken from radio-continuum surveys at 2720,
1420, 820, 408 and 34.5 MHz. We have subtracted the background
radiation, in order to derive the mean brightness temperature of the
SNR alone. First, the temperature of the loop plus background was
determined. For every bin in survey we have some value for
temperature $T_{sum}$. We take all bins within the loop border and
obtained average temperature $<T_{sum}>$ of the loop plus
background. Then, the background alone near the loop was estimated.
We determine temperature of the background near the outer border of
the loop (average value from all data near the outer border of the
loop) and that is the background temperature $T_{back}$. And
finally, the difference of these values is calculated to obtain the
average temperature of the loop at each frequency: average
temperature of the loop is $<T_{sum}> - T_{back}$. The areas over
which an average brightness temperature is determined at each of the
five frequencies are taken to be as similar as possible within the
limits of measurement accuracy. $T_\mathrm{min}$ from Table
\ref{tab01} means the lower temperature limit between the background
and the loop, and $T_\mathrm{max}$ means the upper temperature limit
of the loop. So we used all measured values between $T_\mathrm{min}$
and $T_\mathrm{max}$, inside the corresponding regions of $l$ and
$b$, to calculate the brightness temperature of a loop including the
background. The mean brightness temperature for the loop is found by
subtracting the mean value of background brightness temperature from
the mean value of the brightness temperature over the area of the
loop.

After deriving the mean brightness temperatures
$T_{\mathrm{b},\nu}$, we have converted these values into surface
brightness $\Sigma_{\nu}$ by:

\begin{equation}
\Sigma_\nu = \left( {2k\nu ^2 /c^2 } \right) T_{\mathrm{b},\nu},
\label{equ01}
\end{equation}

\noindent where $k$ is Boltzmann constant and $c$ the speed of
light. Results are given in Table \ref{tab01}.

\begin{table*}[ht!]
\centering
\caption{Temperatures and brightnesses of Cygnus radio
loop at 2720, 1420, 820, 408 and 34.5 MH\lowercase{z}}
\begin{tabular}{c|c|c|c}
\hline
Frequency & Temperature limits & Temperature & Brightness \\
(MHz) & $T_\mathrm{min}$, $T_\mathrm{max}$ (K) & (K) & (10$^{-22}$ W/(m$^2$ Hz Sr)) \\
\hline
2720 & 0.395, 0.66 & 0.160 $\pm$ 0.005 & 3.64 $\pm $ 0.12 \\
1420 & 4.2, 4.9 & 0.49 $\pm$ 0.05 & 3.04 $\pm $ 0.30 \\
820 & 10.1, 14.0 & 2.30 $\pm$ 0.20 & 4.75 $\pm$ 0.40 \\
408 & 41, 72 & 15.3 $\pm$ 1.0 & 7.83 $\pm$ 0.50 \\
34.5 & 34500, 70000 & 13960 $\pm$ 700 & 50.97 $\pm$ 2.56 \\
\hline
\end{tabular}
\label{tab01}
\end{table*}

\section{Results}
\label{sec03}

The radio continuum maps are used for determining the area of the
Cygnus loop and for deriving brightness temperatures over it. At
each of the five frequencies, the areas are determined to be as
similar as possible within the limits of measurement accuracy. There
are still some differences between these areas and we think that the
major causes of differing borders between the five frequencies are
small random and systematic errors in the data. The surface
brightnesses of SNRs must be above the sensitivity limit of the
observations and must be clearly distinguishable from the Galactic
background emission (\citet{gree91}). As it is very difficult to
resolve the fainter parts of the loop from the background, they are
not taken into account. For evaluation brightness temperatures over
the area of the loop we had to take into account background
radiation (see \citet{webs74}). Borders enclosing the spurs are
defined to separate the spur and its background. For the method of
calculation see also \citet{bork07}, \citet{bork08} and
\citet{bork09b}. As mentioned in these papers, if the value of
$T_\mathrm{min}$ is changed by a small amount, the brightness
contours become significantly different. If $T_\mathrm{min}$ is too
small, the area of the spur becomes confused with the background and
it becomes obvious that the border has been incorrectly chosen.

The results are given in Tables \ref{tab01}, \ref{tab02} and
\ref{tab03}. $T_\mathrm{min}$, given in the second column of Table
\ref{tab01}, is the lower temperature limit, while $T_\mathrm{max}$
is the upper temperature of the loop and it is also upper
temperature limit (because there are no other superposed sources).
These temperature limits enable us to distinguish the loop from
background. Then we derived the surface brightnesses using equation
(\ref{equ01}) for each frequency.

The values for brightnesses in $\mathrm{10^{-22}\, W/(m^2\, Hz\,
Sr)}$ can be compared with results for flux densities in Jy. The
flux densities can be transformed into brightnesses or vice versa,
taking into account that the frequencies have to be the same.
Knowing the loop size $\Omega$, the flux densities $S_\nu$ given in
Jy can be transformed to brightnesses given in $\mathrm{10^{-22}\,
W/(m^2\, Hz\, Sr)}$ by:

\begin{equation}
\Sigma_\nu = S_\nu \times 10^{-26} / \Omega.
\label{equ02}
\end{equation}

By use of the spectral indices, the brightnesses can be reduced to
1000 MHz according to relation:

\begin{equation}
\Sigma_{1000} = \Sigma_\nu (1000 / \nu)^{(2 - \beta)},
\label{equ03}
\end{equation}

\noindent where the temperature spectral index $\beta = \alpha + 2$,
and $\alpha$ is the spectral index defined by $S_\nu\propto
\nu^{-\alpha}$.

\subsection*{Spectrum}
\label{sec03a}

\begin{figure}[ht!]
\centering
\includegraphics[width=0.45\textwidth]{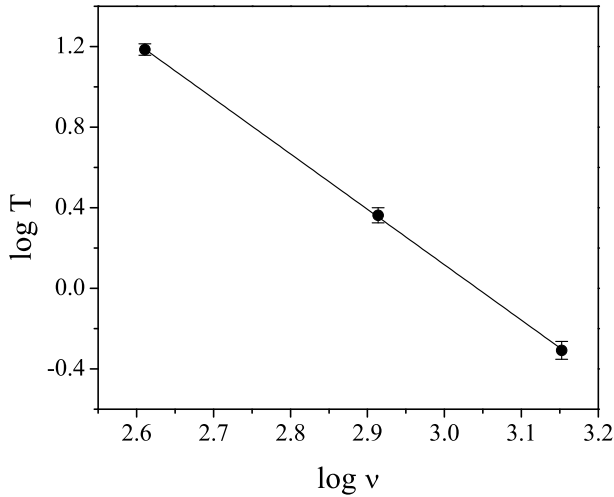}
\caption{Cygnus loop spectrum: temperature versus frequency, for
three measurements -- at 408, 820 and 1420 MHz.}
\label{fig07}
\end{figure}

\begin{figure}[ht!]
\centering
\includegraphics[width=0.45\textwidth]{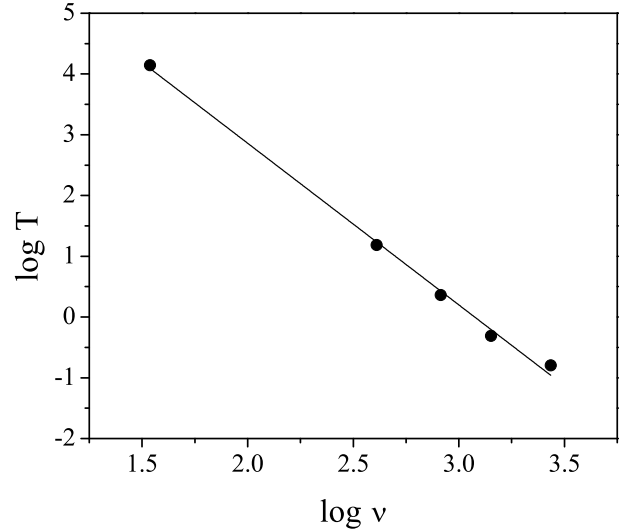}
\caption{Cygnus loop spectrum: temperature versus
frequency, for five measurements -- at 34.5, 408, 820, 1420 and 2720
MHz.}
\label{fig08}
\end{figure}

The spectrum was generated using mean temperatures at five different
frequencies. Best-fit straight line spectrum enables calculation of
spectral index as negative value of the line's direction
coefficient. In Figure \ref{fig07} we give spectrum for the three
middle frequencies: 1420, 820 and 408 MHz because their spectrum is
very well fitted with the straight line (see Figure \ref{fig07}).
Frequencies 2720 and 34.5 MHz, lies on very high and on very low
ends of the spectrum, as presented in Fig. \ref{fig08}. If for
calculation of spectral index we take only three frequencies 1420,
820 and 408 MHz, the result would be $\beta_3$ = 2.76 $\pm$ 0.03.
All five frequencies 2720, 1420, 820, 408 and 34.5 MHz, also from
linear fit, give $\beta_5$ = 2.66 $\pm$ 0.09. It is steeper spectral
index in comparison to average value for Galactic SNRs $\beta = 2.5$
(in Green's catalogue (\citet{gree09}) the adopted value for
spectral index is not given because of its variability). Cygnus loop
is relatively old SNR. The shock wave has to be weak for evolved
SNRs. The steeper spectral indices of SNRs should be expected for
the smaller shock wave velocities. It is result obtained from the
diffuse particle acceleration theory (\citet{bell78a,bell78b}).
Additionally, Cygnus loop probably expands in low density
environment. It looks like large Galactic radio loops - evolved SNRs
with the steep spectral indices, immersed in the low density
environment (see \citet{bork07,bork08}, and references therein).

Obtained values $\beta_3$ = 2.76 $\pm$ 0.03 (from three
frequencies), $\beta_5$ = 2.66 $\pm$ 0.09 (from all five
frequencies), are greater than 2.2 and confirm non-thermal origin of
Cygnus loop emission. From Fig. \ref{fig08} it can be noticed that
linear fit is quite satisfactory. The value for the brightness
temperature spectral index of the Cygnus loop is rather steep. This
is at the high end of the spectral index distribution for SNRs as
suggested in \citet{clar76}.

\begin{figure*}[ht!]
\centering
\includegraphics[height=0.93\textheight]{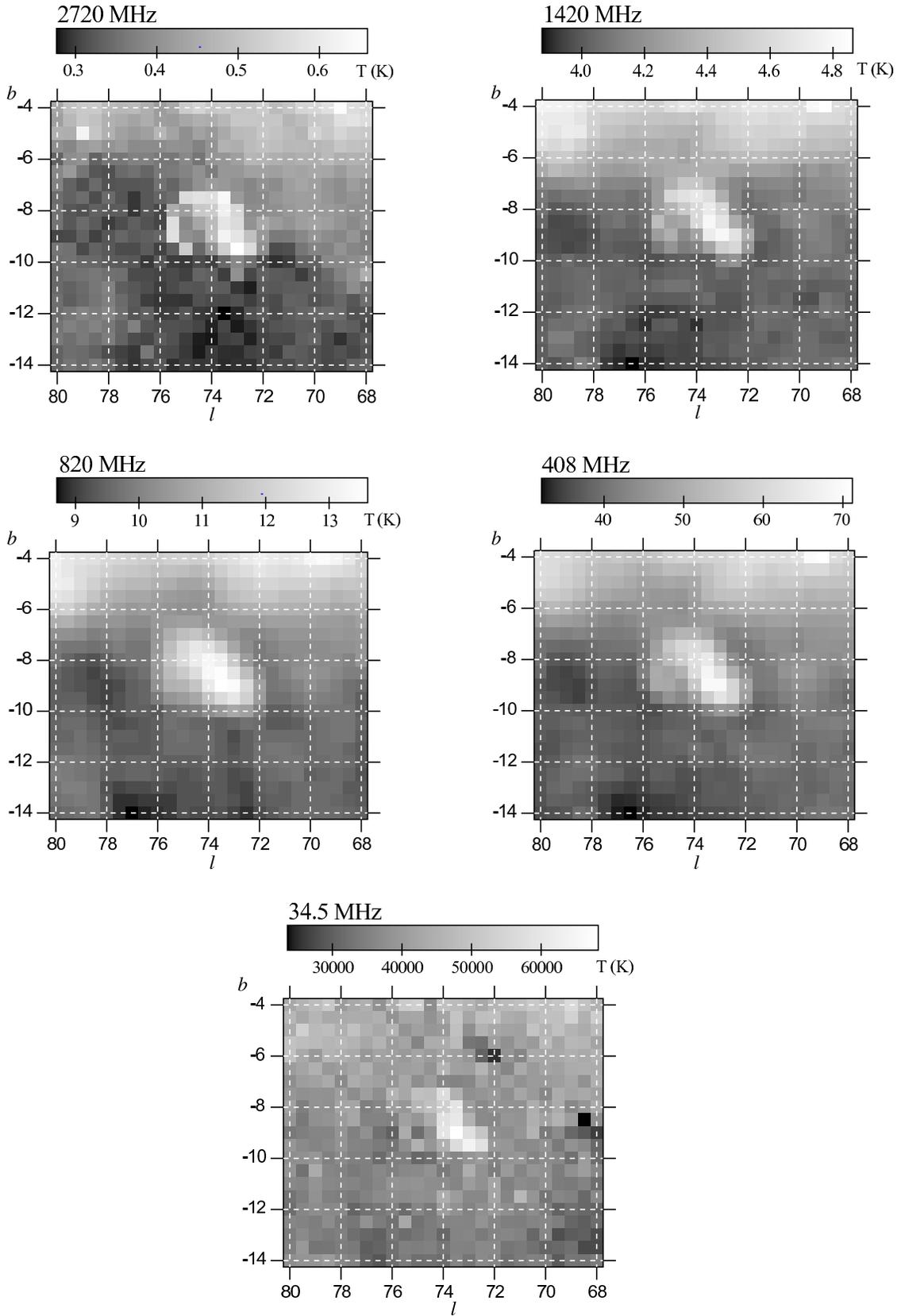}
\caption{The data retabulated to $0^\circ.5 \times 0^\circ.5$
resolution, for the following frequencies: 2720 MHz \emph{(top
left)}, 1420 MHz \emph{(top right)}, 820 MHz \emph{(middle left)},
408 MHz \emph{(middle right)} and 34.5 MHz \emph{(bottom)}. The
HPBWs for these frequencies are 0$^\circ$.35, 0$^\circ$.59,
1$^\circ$.2, 0$^\circ$.85 and 0$^\circ$.7, respectively. The gray
scales of temperatures are given also.}
\label{fig09}
\end{figure*}

\subsection*{$T-T$ plot}
\label{sec03b}

\begin{table*}
\centering
\caption{Spectral indices for Cygnus loop from $T-T$
plots, between 2720, 1420, 820, 408 and 34.5 MH\lowercase{z}}
\begin{tabular}{c|c|c|c|c|c}
\hline
Frequency (MHz) & 2720 & 1420 & 820 & 408 & 34.5 \\
\hline
2720 & / & 1.52 $\pm$ 0.35 & 1.96 $\pm$ 0.53 & 2.35 $\pm$ 0.26 & 2.57 $\pm$ 0.23 \\
1420 & 2.22 $\pm$ 0.76 & / & 2.27 $\pm$ 0.64 & 2.60 $\pm$ 0.28 & 2.67 $\pm$ 0.25 \\
820 & 2.64 $\pm$ 0.61 & 3.10 $\pm$ 0.64 & / & 2.82 $\pm$ 0.31 & 2.82 $\pm$ 0.10 \\
408 & 2.67 $\pm$ 0.26 & 2.79 $\pm$ 0.28 & 3.10 $\pm$ 0.31 & / & 2.78 $\pm$ 0.15 \\
34.5 & 2.86 $\pm$ 0.23 & 2.95 $\pm$ 0.25 & 2.92 $\pm$ 0.10 & 2.94 $\pm$ 0.15 & / \\
\hline
\end{tabular}
\label{tab02}
\end{table*}

The measured data have different resolutions for different
frequencies (see \S 2.1), and therefore in order to obtain $T-T$
plots the data were retabulated so the higher resolution maps are
convolved to the resolution of the lowest resolution map. In that
way we convolved data at 2720, 1420, 408 and 34.5 MHz to $0^\circ.5
\times 0^\circ.5$ resolution, which is the sampling rate of the 820
MHz survey. These retabulated data are presented in Figure
\ref{fig09} for the following frequencies: 2720 MHz (top left), 1420
MHz (top right), 820 MHz (middle left), 408 MHz (middle right) and
34.5 MHz (bottom). Then, for each frequency pair we used only the
common points (with the same $(l,b)$) which belong to the loop area
at both frequencies. In that way we reduced loop area to the same
area for different frequencies. The obtained $T-T$ plots for five
pairs of frequencies enabled calculating the spectral indices. We
calculated two $\beta$ values for each of these pairs: between
2720--1420, 2720--820, 2720--408, 2720--34.5, 1420--820, 1420--408,
1420--34.5, 820--408, 820--34.5 and 408--34.5 MHz and presented it
in Table \ref{tab02}. For each of the ten frequency pairs, by
interchanging the dependent and independent variables we have
obtained two $\beta$ values for each pair and the mean value of
these fit results is adopted as the radio spectral index, as
suggested in \citet{uyan04}. Regarding only three frequencies
(because their spectrum lies on straight line, see Figure
\ref{fig07}) 1420, 820 and 408 MHz, the average value of spectral
index from $T-T$ is $<\beta_{TT}>_3$ = 2.78 $\pm$ 0.41. Taking into
account all five frequencies, we get $<\beta_{TT}>_5$ = 2.63 $\pm$
0.30. It can be noticed that this value agrees well with the
corresponding value obtained from spectrum, as expected (see
\citet{uyan04}).
\newline
\newline

Then we calculated mean value of spectral index: regarding spectrum
and $T-T$ graphs. Between 1420, 820 and 408 MHz we obtained
$<\beta>_3$ = 2.77 $\pm$ 0.22, and between 2720, 1420, 820, 408 and
34.5 MHz we obtained $<\beta>_5$ = 2.64 $\pm$ 0.20.

\section{Discussion}
\label{sec04}

Spectral index variations with position of more spatial features
within the Cygnus loop can be found in paper \citet{leah99}. There
were studied radio spectral indices by $T-T$ plot method between
2695, 1420 and 408 MHz and found that the bright radio filaments all
show negative curvature (steeper at higher frequency), and regions
dominated by diffuse emission show positive curvature (flatter at
higher frequency).

It can be noticed that mean value of spectral index for Cygnus loop
is little higher then the corresponding value obtained from articles
(see \citet{uyan04} or \citet{leah98}). Reason for this is different
areas used for data of Cygnus loop in these papers. In both papers
they used square areas (see page 917 from \citet{uyan04} and page
786 from \citet{leah98}). In these square areas they take into
account parts of Cygnus loop and parts of background radiation near
Cygnus loop (these parts we do not take into account). We take in
account only higher intensity regions from these areas (see Figures
\ref{fig01}--\ref{fig05}).

\begin{table}[hb!]
\centering
\caption{Brightness of Cygnus radio loop reduced to 1000
MH\lowercase{z}, using spectral index derived in this paper:
$\beta_5$ = 2.66 $\pm$ 0.09 (from the spectrum of all five
frequencies 2720, 1420, 820, 408 and 34.5 MH\lowercase{z})}
\begin{tabular}{c|c}
\hline
Frequency & Brightness at 1000 MHz \\
(MHz) & (10$^{-22}$ W/(m$^2$ Hz Sr)) \\
\hline
2720 & 7.05 $\pm$ 0.86 \\
1420 & 3.84 $\pm$ 0.52 \\
820 & 4.17 $\pm$ 0.29 \\
408 & 4.33 $\pm$ 0.07 \\
34.5 & 5.52 $\pm$ 1.40 \\
\hline
\end{tabular}
\label{tab03}
\end{table}

In order to compare our values for brightnesses with results for
fluxes given in other papers, we transform their calculated fluxes
into brightnesses at 1000 MHz. Knowing the loop size $\Omega$ we
reduce the flux densities given in Jy to brightnesses given in
$\mathrm{10^{-22}\, W/(m^2\, Hz\, Sr)}$ by equation (\ref{equ02})
and then by use of the spectral indices, the brightnesses are
extrapolated to 1000 MHz according to relation (\ref{equ03}).

In our previous paper (\citet{bork09a}) we transformed the flux
densities for Cygnus Loop given in \citet{roge99} at 22 MHz ($S_\nu$
= 1378 Jy) and \citet{reic03} at 863 MHz ($S_\nu$ = 184 Jy) into
$\Sigma_{1000}$ using loop size $\Omega = 240' \times 170'$ from
\citet{reic03} and spectral index $\beta = 2.49$ from
\citet{trus02}. We obtained that values from (\citet{bork09a}) agree
with previous data. But now in this paper, we use our derived value
for spectral index $\beta_5 = 2.66 \pm 0.09$, and obtain even better
agreement. From flux densities given in mentioned papers, we
calculated the following values for radiation intensities:
$\Sigma_{1000}$ = $3.22 \times \mathrm{10^{-22}\, W/(m^2\, Hz\,
Sr)}$ from flux given in \citet{roge99} and $\Sigma_{1000}$ = $4.84
\times \mathrm{10^{-22}\, W/(m^2\, Hz\, Sr)}$ from flux given in
\citet{reic03}.

Using our calculated $\beta_5$ and applying relation (\ref{equ03})
to our calculated values of $\Sigma_\nu$ at 2720, 1420, 820, 408 and
34.5 MHz, for $\Sigma_{1000}$ we obtain values given in Table
\ref{tab03}. Absolute errors for brightnesses at 1000 MHz are
calculated in this way: $\Delta \Sigma_{1000} = \left(
{\Sigma_{1000} / \Sigma_{\nu 1}} \right) \left( {\Delta \Sigma_{\nu
1} + \Sigma_{\nu 1} \cdot \ln \left( {\nu_1 / 1000} \right) \cdot
\Delta \beta} \right)$, where $\nu_1$ takes values 2720, 1420, 820,
408 and 34.5 MHz.

\begin{table}[ht!]
\centering \caption{The Cygnus loop areas $\Omega$ and its angular
radii $\theta$ at the five frequencies}
\begin{tabular}{c|c|c}
\hline
Frequency (MHz) & $\Omega$ ($(^\circ)^2$) & $\theta$ ($^\circ$) \\
\hline
2720 & 7.52 & 1.547 \\
1420 & 8.64 & 1.658 \\
820 & 9.46 & 1.735 \\
408 & 9.88 & 1.773 \\
34.5 & 8.44 & 1.639 \\
\hline
\end{tabular}
\label{tab04}
\end{table}

\begin{figure}[ht!]
\centering
\includegraphics[width=0.45\textwidth]{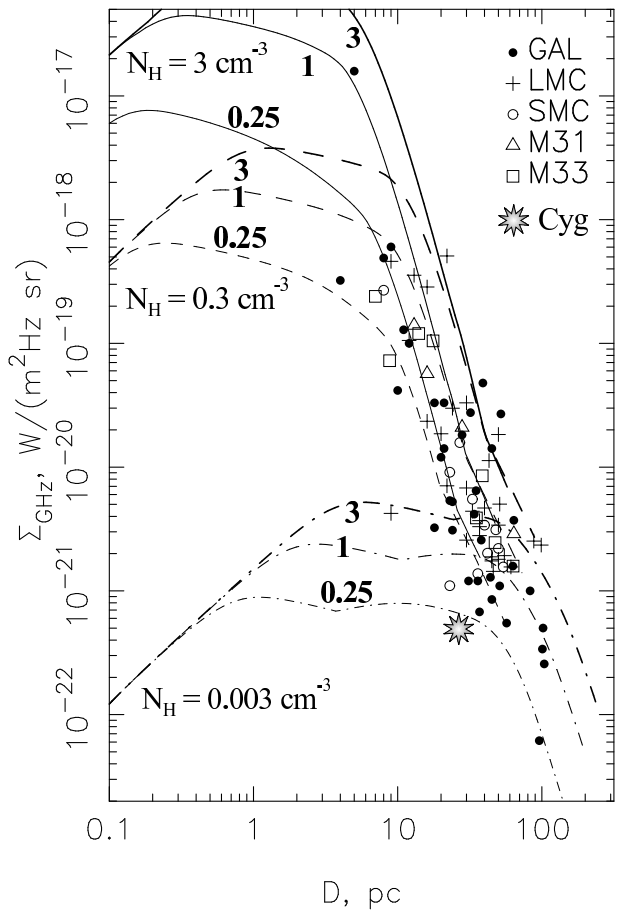}
\caption{The surface brightness to diameter diagram from Berezhko \&
V\"{o}lk (2004), with value for Cygnus loop added. Three different
densities for the interstellar matter (ISM) ($N_\mathrm{H}$ = 3, 0.3
and 0.003 cm$^{-3}$) are presented, plus three values for the (SN)
explosion energy ($E_\mathrm{SN}$ = 0.25, 1 and 3 $\times 10^{51}$ erg).}
\label{fig10}
\end{figure}

Information about variations of brightness and spectrum over radio
image is presented by \citet{leah98} and by \citet{uyan02}. The
average values reduce the information imprinted in variations of
brightness and spectrum over the radio image but give us other
interesting information, like explosion energy, surface brightness
and distance to the Cygnus radio loop.

With regard to the loop's borders in Figs. \ref{fig01}--\ref{fig05},
we derived loop area at each frequency, as well as its corresponding
angular radii (see Table \ref{tab04}). An angular radius is obtained
in this way: our derived loop area, when approximated with the
circle of the same area, gives possibility of determining angular
radius as $\theta$ = $\sqrt{\Omega / \pi}$. These areas can be
compared to the areas calculated by other authors, e.g.
9$(^\circ)^2$.80 at 1420 MHz (\citet{leah02}) and 11$(^\circ)^2$.33
at 863 MHz (\citet{reic03}). When their areas are recalculated into
angular radii in the way we described, the result are: 1$^\circ$.77
at 1420 MHz (\citet{leah02}) and 1$^\circ$.90 at 863 MHz
(\citet{reic03}). It can be noticed that we obtained somewhat
smaller values, but it has to be taken into account that other
authors estimated only the rectangular map size while we estimated
the loop size (inside its contour borders) exactly.

The relation between surface brightness ($\Sigma$) and diameter
($D$) for supernova remnants (SNRs) - so-called the $\Sigma-$D
relation - is appropriate for description of the radio brightness
evolution of these sources. The empirical relations should be used
with caution because of the limited usefulness due to selection
effects (see \citet{gree09}, \citet{uros10} and references therein).
The updated theoretical relations were derived by \citet{duri86},
based on the Bell's (\cite{bell78a,bell78b}) diffuse shock particle
acceleration theory, and by Berezhko and V\"{o}lk (\citet{bere04})
based on the non-linear kinetic theory of the diffuse shock
acceleration mechanism. The $\Sigma-D$ diagram at 1 GHz with the
theoretically derived evolutionary tracks taken from \citet{bere04}
with our value for the Cygnus loop superposed, is shown in Fig.
\ref{fig10}. The $\Sigma-D$ diagram at 1 GHz taken from
\citet{bere04} with the derived value for the Cygnus loop
superposed, is shown in Fig. \ref{fig10}. With aim to superpose
Cygnus position, we used the following results of our calculations:
the mean value of brightnesses at 1 GHz and the diameter, calculated
as $D = 2 r \sin \bar \theta$, where $\bar \theta$ = 1$^\circ$.67
being mean angular radius for all five frequencies. The distance $r$
= 0.44 kpc is taken from Green's catalogue of SNRs (\citet{gree09}).
So we added this point to the diagram: ($D$, $\Sigma$) = (25.7 pc,
4.98 $\times$ 10$^{-22}$ W/(m$^2$ Hz Sr)). From its position in this
diagram it can be concluded that Cygnus loop evolves in the low
density environment and the initial energy of supernova (SN)
explosion was relatively low (see Fig. \ref{fig10}).

\subsection*{Flux density spectrum}
\label{sec04a}

\begin{figure}[ht!]
\centering
\includegraphics[width=0.45\textwidth]{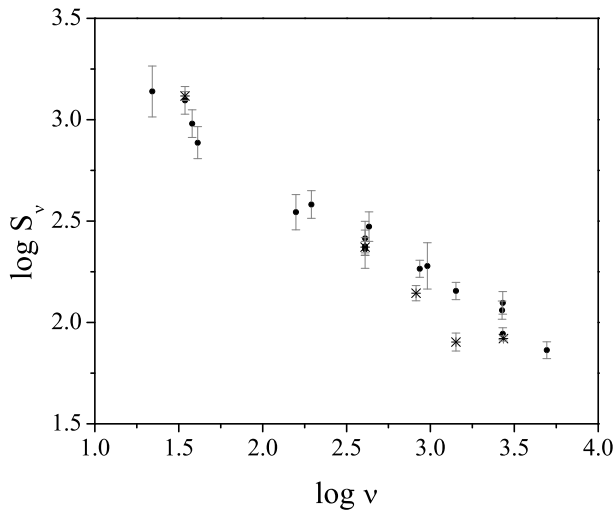}
\caption{Spectrum of Cygnus Loop: flux versus frequency for more different frequencies,
obtained from all available flux density values (values are given in Table 3 in Uyaniker et al. (2004))
and from values calculated in this paper. We identify our values by asterisks.}
\label{fig11}
\end{figure}

On the basis of values for brightnesses given in the fourth column
of Table \ref{tab01} and our calculated values for loop size
$\Omega$ (Table \ref{tab04}), we derived flux values in Jy. The
calculated flux densities we give in Table \ref{tab05}. As the
Cygnus loop is well studied SNR, it is possible to perform a
multi-frequency spectral study. In the Fig. \ref{fig11} we present a
summary of some results obtained from other papers and from our
study. Our results are labeled with asterisks. So, in this figure,
we have seventeen results for the flux density values $S_{\nu}$ of
Cygnus loop derived in several papers (values are given in Table 3
of \citet{uyan04} and five results from our paper.

It can be seen from Fig. \ref{fig11} that our added flux values fit
very well among other fluxes, which shows correctness of our method,
and also that we determined loop area and its brightness well.

\begin{table}[ht!]
\centering \caption{Flux densities (J\lowercase{y}) that we
calculated at the five frequencies}
\begin{tabular}{c|c}
\hline
Frequency (MHz) & Flux density (Jy) \\
\hline
2720 & 83.37 $\pm$ 2.60 \\
1420 & 80.05 $\pm$ 8.13 \\
820 & 139.35 $\pm$ 12.10 \\
408 & 235.44 $\pm$ 15.36 \\
34.5 & 1309.04 $\pm$ 65.65 \\
\hline
\end{tabular}
\label{tab05}
\end{table}

\section{Conclusions}
\label{sec06}

The main result is method for determination brightness temperature
given in article \citet{bork07} which is developed for large radio
loops, and we show that this method is very good for much smaller
loop, e.g. Cygnus loop. We check the method by applying it to the
Cygnus radio loop at the wide range of frequencies. It is in good
agrement with method of $T-T$ graphics and results obtained with
another methods. This method is quite simple because we use
brightness temperature isoline to define border of a Cygnus loop.
Other authors are using different squared or rectangular areas to
determine area of the loop and calculate spectral indices,
brightness temperature and the flux density of the loop
(\citet{uyan04}, page 917 and \citet{leah98}, page 786). Also, we
calculated flux for Cygnus loop and compare our results with results
of other authors in different ranges of frequencies. We show that
our results are in good agrement with these results.

We estimated the temperatures and brightness of the Cygnus loop SNR
on the basis of observations of the continuum radio emission at the
frequencies: 2720, 1420, 820, 408 and 34.5 MHz (this paper and
\citet{bork09a}). The sensitivity of the brightness temperatures
are: 5 mK for 2720 MHz, 50 mK for 1420 MHz, 0.2 K for 820 MHz, 1.0 K
for 408 MHz and about 700 K $T_\mathrm{b}$ for 34.5 MHz. At the
frequency of 2720 MHz the measurements are the most precise (with
the least relative errors) so positions of the brightness
temperature contours of the loop are the most realistic for this
frequency.

Borders between the five frequencies are somewhat different for this
SNR probably due to small, random and systematic errors in the
calibrated data. Also, we suppose there are uncertainties of about
($2 \times \Delta T$) 10 mK for 2720 MHz, 100 mK for 1420 MHz, 0.4 K
for 820 MHz, 2.0 K for 408 MHz and about 1400 K $T_\mathrm{b}$ for
34.5 MHz, in the border due to measurement errors, and there is a
tiny difference in the absorption of radio emission in the
interstellar medium at different wavelengths (\citet{pach70}).

We determined average brightness temperature from a Cygnus radio
loop region, after subtraction of a background level. Our obtained
values (when all reduced to 1000 MHz for comparison) are in good
agreement with the earlier results. We present the radio continuum
spectrum of the Cygnus loop using average brightness temperatures at
five different frequencies. As it can be seen from Figs. \ref{fig07}
and \ref{fig08}, given linear fit provides reliable spectral index.
We present the $T-T$ plots which enable the calculation of spectral
index, too.

Also, from our results can be concluded that Cygnus loop evolves in
the low density environment and the initial energy of SN explosion
was relatively low. This can be read after superposing the position
of this loop to theoretical $\Sigma-D$ diagram form \citet{bere04},
which was derived from non-linear kinetic theory of diffuse shock
acceleration mechanism.

We showed that method for defining a loop border and for determining
the values of temperature and brightness, which we developed for
main Galactic loops I-VI, could be applicable to all SNRs.

\begin{acknowledgments}
The authors are grateful to the referee whose suggestions
substantially improved the paper. This research is supported by the
Ministry of Science of the Republic of Serbia through project No.
176005.
\end{acknowledgments}


\begin{thebibliography}{00}   

\bibitem[McKee \& Ostriker (1977)]{mcke77} McKee, C. F. \& Ostriker,
    J. P. 1977, ApJ, 218, 148

\bibitem[Salter (1983)]{salt83} Salter, C. J. 1983, Bull. Astron.
    Soc. India, 11, 1

\bibitem[Green (2004)]{gree04} Green, D. A. 2004, Bull. Astron. Soc.
    India, 32, 335

\bibitem[Green (2006)]{gree06} Green, D. A. 2006, A Catalogue of
    Galactic Supernova Remnants (2006 April version), Cavendish
    Laboratory, Cambridge, UK

\bibitem[Green (2009)]{gree09} Green, D. A. 2009, Bull. Astron.
    Soc. India, 37, 45

\bibitem[Green (1984)]{gree84} Green, D. A. 1984, MNRAS, 211, 433

\bibitem[Leahy, Roger, \& Ballantyne (1997)]{leah97} Leahy, D. A.,
    Roger, R. S. \& Ballantyne, D. 1997, AJ, 114, 2081L

\bibitem[Aschenbach \& Leahy (1999)]{asch99} Aschenbach, B., \&
    Leahy, D. A. 1999, A\&A, 341, 602

\bibitem[Uyaniker et al. (2002)]{uyan02} Uyaniker, B., Reich, W.,
    Yar, A., Kothes, R., \& F\"{u}rst, E. 2002, A\&A, 389, L61

\bibitem[Patnaude et al. (2002)]{patn02} Patnaude, D. J., Fesen, R.
    A., Raymond, J. C., Levenson, N. A., Graham, J. R., \& Wallace,
    D. J. 2002, AJ, 124, 2118

\bibitem[Uyaniker et al. (2004)]{uyan04} Uyaniker, B., Reich, W.,
    Yar, A., \& F\"{u}rst, E. 2004, A\&A, 426, 909

\bibitem[Leahy (2002)]{leah02} Leahy, D. A. 2002, AJ, 123, 2689

\bibitem[Sun et al. (2006)]{sun06} Sun, X. H., Reich, W., Han, J.
    L., Reich, P., \& Wielebinski, R. 2006, A\&A, 447, 937

\bibitem[Kundu \& Becker (1972)]{kund72} Kundu, M. R. \& Becker, R.
    H. 1972, AJ, 77, 459

\bibitem[Keen et al. (1973)]{keen73} Keen, N. J., Wilson, W. E.,
    Haslam, C. G. T., Graham, D. A. \& Thomasson, P. 1973, A\&A, 28,
    197

\bibitem[Dickel \& Willis (1980)]{dick80} Dickel, J. R.
    \& Willis, A. G. 1980, A\&A, 85, 55

\bibitem[Reif et al. (1987)]{reif87} Reif, K., Reich, W., Steffen,
    P., M\"{u}ller, P., \& Weiland, H. 1987, Mitt. Astr. Ges., 70,
    419

\bibitem[Reich \& Reich (1986)]{reic86} Reich, P. \& Reich, W.
    1986 A\&AS, 63, 205

\bibitem[Berkhuijsen (1972)]{berk72} Berkhuijsen, E. M. 1972, A\&AS,
    5, 263

\bibitem[Haslam et al. (1982)]{hasl82} Haslam, C. G. T., Salter, C.
    J., Stoffel, H., \& Wilson, W. E. 1982, A\&AS, 47, 1

\bibitem[Dwarakanath \& Udaya Shankar (1990)]{dwar90}
    Dwarakanath, K. S. \& Udaya Shankar, N. 1990, J. Astrophys. Astr., 11, 323

\bibitem[Borka Jovanovi\'{c} \& Uro\v{s}evi\'{c} (2009a)]{bork09a}
    Borka Jovanovi\'{c}, V. \& Uro\v{s}evi\'{c}, D. 2009a, Publ.
    Astron. Obs. Belgrade, 86, 101

\bibitem[Reich, Zhang \& F\"{u}rst (2003)]{reic03} Reich, W., Zhang,
    X., \& F\"{u}rst, E. 2003, A\&A, 408, 961

\bibitem[Roger et al. (1999)]{roge99} Roger, R. S., Costain, C. H.,
    Landecker, T. L., \& Swerdlyk, C. M. 1999, A\&AS, 137, 7

\bibitem[Borka (2007)]{bork07} Borka, V. 2007, MNRAS, 376, 634

\bibitem[Leahy \& Roger (1998)]{leah98} Leahy, D. A. \&
    Roger, R. S. 1998, ApJ, 505, 784

\bibitem[Borka, Milogradov-Turin, \& Uro\v{s}evi\'{c}
    (2008)]{bork08} Borka, V., Milogradov-Turin, J., \&
    Uro\v{s}evi\'{c}, D. 2008, Astron. Nachr., 329, 397

\bibitem[Borka Jovanovi\'{c} \& Uro\v{s}evi\'{c} (2009b)]{bork09b}
    Borka Jovanovi\'{c}, V. \& Uro\v{s}evi\'{c}, D. 2009b, Astron.
    Nachr., 330, 741

\bibitem[Green (1991)]{gree91} Green, D. A. 1991, PASP, 103, 209

\bibitem[Webster (1974)]{webs74} Webster, A. S. 1974, MNRAS, 166,
    355

\bibitem[Bell (1978a)]{bell78a} Bell, A. R. 1978a, MNRAS, 182,
    147

\bibitem[Bell (1978b)]{bell78b} Bell, A. R. 1978b, MNRAS, 182,
    443

\bibitem[Clark \& Caswell (1976)]{clar76} Clark, D. H. \& Caswell,
    J. L. 1976, MNRAS, 174, 267

\bibitem[Leahy (1999)]{leah99} Leahy, D. A. 1999, ASP Conferece
    Series, 168, 437L

\bibitem[Trushkin (2002)]{trus02} Trushkin, S. A. 2002, CATS
    Database - Astrophysical CATalogs support System, SNRs Spectra
    Request Form, \url{http://www.sao.ru/cats/snr_spectra.html}

\bibitem[Uro\v{s}evi\'{c} et al. (2010)]{uros10} Uro\v{s}evi\'{c},
    D., Vukoti\'{c}, B, Arbutina, B., and Sarevska, M. 2010, ApJ,
    719, 950

\bibitem[Duric \& Seaquist (1986)]{duri86} Duric, N., \&
    Seaquist, E. R. 1986, ApJ, 301, 308

\bibitem[Berezhko \& V\"{o}lk (2004)]{bere04} Berezhko, E. G., \&
    V\"{o}lk, H. J. 2004, A\&A 427, 525

\bibitem[Pacholczyk (1970)]{pach70} Pacholczyk, A. G. 1970, Radio
    Astrophysics, Freeman W. H. \& Company, San Francisco

\end{thebibliography}
\end{document}